\newcommand{\beq}{\begin{equation}}
\newcommand{\eeq}{\end{equation}}
\newcommand{\bea}{\begin{eqnarray}}
\newcommand{\eea}{\end{eqnarray}}
\newcommand{\bear}{\begin{eqnarray*}}
\newcommand{\eear}{\end{eqnarray*}}

\newcommand{\rf}[1]{(\ref{#1})}
\documentstyle[aps,preprint]{revtex}
\begin{document}

\draft

\title
{Exact Solution of Asymmetric Diffusion With Second-Class Particles of Arbitrary Size}

\author{
F. C. Alcaraz}  
\address{Departamento de F\'{\i}sica, 
Universidade Federal de S\~ao Carlos, 13565-905, S\~ao Carlos, SP
Brazil}
\author{ R. Z. Bariev}

\address{Departamento de F\'{\i}sica, 
Universidade Federal de S\~ao Carlos, 13565-905, S\~ao Carlos, SP
Brazil}

\address{The Kazan Physico-Technical Institute of the Russian 
Academy of Sciences, 
Kazan 420029, Russia}

\maketitle

\begin{abstract}

The exact solution of the asymmetric exclusion problem with 
first- and scond-class particles 
is presented. In this model the particles 
(size 1) of both classes are located at lattice points, and diffuse with  
equal asymmetric rates, but particles in the first class do not distinguish 
those in the second class from holes (empty sites). We generalize and 
solve exactly this model by considering molecules in the first and 
second class with sizes $s_1$ and $s_2$ ($s_1,s_2 = 0,1,2,\ldots$), in  
units of lattice spacing, respectively.	The solution is derived by 
a Bethe ansatz of nested type. We give a simple pedagogical 
presentation of the Bethe ansatz solution of the problem which 
can easily be followed by a reader not specialized in exactly integrable  
models.

\end{abstract}


\narrowtext    
\section{Introduction}

	The similarity between the master equation describing time 
fluctuations in nonequilibrium problems and the Schr\"odinger equation describing the 
quantum fluctuations of quantum spin chains turns out to be fruitful 
for both areas of research \cite{lushi} - \cite{alc-bar1}. 
Since many quantum chains are known to be 
exactly integrable through the Bethe ansatz, this provides exact information 
on the related stochastic model. At the same time the classical physical  
intuition and probabilistic methods successfully applied to nonequilibrium 
systems give new insights into the understanding of the physical and 
algebraic properties of quantum chains.

An example of this fruitful interchange is the problem of 
asymmetric diffusion of hard-core particles on the one dimensional lattice. 
This model 
is related to the exactly integrable anisotropic Heisenberg 
chain \cite{yang} 
(XXZ model). However if we demand this quantum chain to be invariant 
under a quantum group symmetry $U_q(SU(2))$, we have to introduce, for the 
equilibrium statistical system, unphysical surface terms, which on the other 
hand have a nice and simple interpretation for the related nonequilibrium 
stochastic system \cite{alcrit1,alcrit2}.

In the area of exactly integrable models it is well known that one of the 
possible extensions of the spin-$\frac{1}{2}$ XXZ chain to higher spins, is 
the anisotropic spin-1 Sutherland model (grading $\epsilon_1 = \epsilon_2 = 
\epsilon_3 = 1$) \cite{sutherland}. 
On the other hand in the area of diffusion limited 
reactions a simple extension of the asymmetric diffusion problem is the 
problem of diffusion with first and second class of particles\cite{bold} - 
\cite{ferra2} . In this 
problem a mixture of two classes of hard-core particles 
diffuses on the lattice. Particles belonging to the first class ignore 
the presence of those in the second class, i. e., they see them in the 
same way as they see the holes (empty sites). In \cite{alcrit1} 
it was shown that for 
open boundary conditions the anisotropic spin-1 Sutherland model and this 
last stochastic model are exactly related, the Hamiltonian governing 
the quantum or time fluctuations of both models being given in terms of generators 
of a Hecke algebra, invariant under the quantum group $U_qSU(3)$. In this 
paper we are going to derive the exact solution of the associated quantum 
chain, in a closed lattice. Recently \cite{alc-bar1} (see also \cite{sasawada})
we have shown that without losing 
its exact integrability, we can consider the problem of asymmetric 
diffusion with an arbitrary mixture of molecules 
with different sizes (even zero),  as 
long they do not interchange positions, that is, there is no reactions. 
Motivated by these results we are going to extend the asymmetric diffusion 
problem with a second class of particles, to the case where the particles in 
each class have an arbitrary size, in units of the lattice spacing. Unlike
the case of asymmetric diffusion problem, we have in this case a 
nested Bethe ansatz \cite{takh}. 

The paper is organized as follows. In the next section we introduce the 
generalized asymmetric model with second-class particles and derive the 
associated quantum chain. In section 3 the Bethe ansatz solution of the 
model is presented in a pedagogical and self contained way, which should  
be simple to follow by an audience not specialized in exactly integrable  
models. Finally in section 4 
we present our conclusions, with some possible  generalizations of the 
stochastic problem considered in this paper, and some perspectives on
future work.

\section{The generalized asymmetric diffusion model
with first- and second-class of particles}

	A simple extension of the asymmetric exclusion model, where hard-core 
particles diffuse on the lattice, is the problem where a mixture of 
particles belonging to different classes (first and second class) diffuses 
on the lattice. This problem was used to describe shocks \cite{bold} -
\cite{ferra2}  in nonequilibrium 
and also has a stationary probability distribution that can be expressed  
via the matrix-product ansatz \cite{derrida1}. 
In this model we have $n_1$ and $n_2$ 
molecules belonging to the first and second class, respectively. Both classes 
of molecules diffuse asymmetrically, but with the same asymmetrical rates, 
whenever they encounter empty sites (holes) at nearest-neighbor sites. However,
when molecules of different classes are at their minimum separation, the  
molecules of the first class exchange position with the same rate as they 
diffuse, and consequently the first-class molecules see no difference 
between molecules belonging to the second class and holes. 

We now introduce a generalization of the above model, where instead of having
unit size, the molecules in the first and second class 
have in general distinct sizes $s_1$ and $s_2$ ($s_1,s_2 =1,2,\ldots$), respectively,
in units of lattice spacing. In Fig. 1
we show some examples of molecules of different sizes. We may think of a  
molecule of size $s$ as  formed by $s$ monomers (size 1), and for 
simplicity, we define the position of the molecule as the center 
of its leftmost monomer. The molecules have a hard-core repulsion: the 
minimum distance $d_{\alpha \beta}$, in units of the lattice spacing, between 
molecules $\alpha$ and $\beta$, with $\alpha$ in the left, is given by 
$d_{\alpha\beta} =s_{\alpha}$.  In order to describe the occupancy of a given 
configuration of molecules we define, at each site $i$ of a lattice with  
$N$ sites, a variable $\beta_i$ ($i=1,2,\ldots,N$), taking the values 
$\beta_i = 0,1$ and 2, representing site $i$ empty (size $s_0 =1$), occupied by
a molecule of class 1 (size $s_1$) or a molecule of class 2 (size $s_2$), 
respectively. Then the allowed configurations are given by the set 
$\{\beta_i\}$ ($i=1,\ldots,N$), where for each pair 
$(\beta_i,\beta_j)\ne 0$ with $j>i$ we should have $j -i \geq s_{\beta_i}$.

The time evolution of the probability distribution 
$ P(\lbrace\beta\rbrace,t)$, 
of a given configuration $\{\beta\}$ is given by the master equation 
\beq
\label{1} 
\frac{\partial  P(\lbrace\beta\rbrace,t)}{\partial t} = 
- \Gamma(\lbrace\beta\rbrace 
\rightarrow \lbrace\beta'\rbrace)  P(\lbrace\beta\rbrace,t) +
\Gamma(\lbrace\beta'\rbrace 
\rightarrow \lbrace\beta\rbrace)  P(\lbrace\beta'\rbrace,t),
\eeq
where $\Gamma(\lbrace\beta\rbrace \rightarrow \lbrace\beta'\rbrace)$ is 
the transition rate for configuration $\lbrace\beta\rbrace$ to change to
$\lbrace\beta'\rbrace$.  In the present model we only allow,
whenever it is possible, the particles  to diffuse to  
nearest-neighbor sites, or to exchange positions. 
The possible motions are diffusion to the right
\bea
\label{2}
1_i\;\;\emptyset_{i+s_1} &\rightarrow& \emptyset_i\;\;1_{i+1},\;\;\;\;\;\;\;\;
\;\;\;\;\;\;\;(\mbox {rate} \;\; \Gamma_R)\nonumber\\
2_i\;\;\emptyset_{i+s_2} &\rightarrow& \emptyset_i\;\;2_{i+1},\;\;\;\;\;\;\;\;
\;\;\;\;\;\;\;(\mbox {rate} \;\; \Gamma_R),
\eea
diffusion to the left 
\bea
\label{3}
\emptyset_i\;\;1_{i+1} &\rightarrow& 1_i\;\;\emptyset_{i+s_1},\;\;\;\;\;\;\;\;
\;\;\;\;\;\;\;(\mbox {rate} \;\; \Gamma_L)\nonumber\\
\emptyset_i\;\;2_{i+1} &\rightarrow& 2_i\;\;\emptyset_{i+s_2},\;\;\;\;\;\;\;\;
\;\;\;\;\;\;\;(\mbox {rate} \;\; \Gamma_L),
\eea
and interchange of particles
\bea 
\label{4}
1_i\;\;2_{i+s_1} &\rightarrow& 2_i\;\;1_{i+s_2},\;\;\;\;\;\;\;\;
\;\;\;\;\;\;\;(\mbox {rate}\;\;  \Gamma_R)\nonumber\\
2_i\;\;1_{i+s_2} &\rightarrow& 1_i\;\;2_{i+s_1},\;\;\;\;\;\;\;\;
\;\;\;\;\;\;\;(\mbox {rate} \;\; \Gamma_L).
\eea
As we see from \rf{4}, particles in the first class interchange positions with 
those of second class with the same rate as they interchange positions with 
the empty sites (diffusion). We should remark however that unless the second 
class  particles have unit size ($s_2 =1$), the net effect of the second 
class particles in those of the first class is distinct from the effect 
produced by the holes, since as the result of the exchange the first class 
of particles will move by $s_2$ lattice size units, accelerating its diffusion.

 The master equation \rf{1} can be written as 
a Schr\"odinger equation in Euclidean time (see Ref. \cite{alcrit1} for general
application for two body processes)
\beq
\label{5}
\frac{\partial| P>}{\partial t} = -H | P>,
\eeq
if we interpret $| P> \equiv  P(\lbrace\beta\rbrace, t)$ 
as the associated wave 
function. If we represent $\beta_i$ as $|\beta>_i$ the vector 
$|\beta>_1\otimes |\beta>_2 \otimes \cdots \otimes |\beta>_N$ will give us 
the associated 
Hilbert space. The process \rf{2}-\rf{4} gives us the Hamiltonian (see 
Ref. \cite{alcrit1} for general applications)
\bea
\label{6}
H &=& D \sum_{j} H_{j}  \nonumber \\
H_j &=& - {\cal P} \{ \sum_{\alpha=1}^2 \left[ \epsilon_+(E_{j}^{0\alpha}
E_{j+1}^{\alpha 0} - E_j^{\alpha \alpha} E_{j+1}^{0 0}) +
\epsilon_-(E_j^{\alpha 0}E_{j+1}^{0\alpha} - E_j^{0 0} E_{j+1}^{\alpha \alpha}) 
\right] \nonumber \\
&& + \epsilon_+(E_j^{21}E_{j+s_2}^{10}E_{j+s_1}^{02} - E_j^{11}E_{j+s_2}^{00}
E_{j+s_1}^{22}) + 
 \epsilon_-(E_j^{12}E_{j+s_1}^{20}E_{j+s_2}^{01} - E_j^{22}E_{j+s_1}^{00}
E_{j+s_2}^{11})\} {\cal P}
\eea
with 
\beq
\label{7}
D = \Gamma_R + \Gamma_L,\;\;\; \epsilon_+ =\frac{\Gamma_R}{\Gamma_R + 
\Gamma_L},\;\;\; \epsilon_- = \frac{\Gamma_L}{\Gamma_R + \Gamma_L} \quad 
(\epsilon_+ + \epsilon_- = 1),
\eeq
and periodic boundary conditions. The matrices $E^{\alpha,\beta}$ are $3\times 
3$ matrices with a single nonzero
element $(E^{\alpha,\beta})_{i,j} = \delta_{\alpha,i}\delta_{\beta,j}$  
($\alpha,\beta,i,j \in {\rm Z})$. The projector ${\cal P}$ in \rf{6}, 
projects out from the associated  Hilbert 
space the vectors  $|\lbrace\beta\rbrace>$ which represent forbidden 
positions of the molecules due to their finite size, which mathematically 
 means that for all $i,j$ with
$\beta_i,\beta_j \ne 0,\;\; |i - j| \ge  s_{\beta_i} (j >i)$.
The constant D in  \rf{6} fixes the time scale and for simplicity we
 chose $D = 1$. A particular simplification of \rf{6}
occurs when  molecules of the first and second class have the same  size  
$s_1 =s_2 =s$. In this case
the Hamiltonian can be expressed as an anisotropic nearest-neighbor 
interaction spin-1 $SU(3)$ chain. Moreover in the case where their  
sizes are unity  ($s=1$) the model  may be related to the $SU(3)$ Sutherland 
model with twisted boundary conditions \cite{sutherland}.

\section{ The Bethe ansatz equations}

We present in this section the exact solution of the general quantum chain 
\rf{6}. 
Since the present paper is going to appear in a special issue where the main 
subject is nonequilibrium physics, we are going to present a pedagogical and
self-contained derivation of the exact solution, that can clearly be followed 
by a nonexpert in the arena of exactly solvable models in statistical 
mechanics. 

Due to the conservation of particles in the diffusion and interchange 
processes the 
total number of particles $n_1$ and $n_2$ of particles of class 1 and 2 
are good quantum numbers and consequently we can split 
the associated Hilbert space into block disjoint sectors labeled by the 
numbers $n_1$ and $n_2$ ($n_1 = 0,1,\ldots,n; \; n_2 =n-n_1;\; n=n_1+n_2$).
We therefore consider the eigenvalue equation 
\beq
\label{8}
H |n_1,n_2> = E|n_1,n_2>,
\eeq
where
\beq
\label{9}
|n_1,n_2> = \sum_{\{Q\}} \sum_{\{x\}}f(x_1,Q_1;\ldots;x_n,Q_n)
|x_1,Q_1;\ldots;x_n,Q_n>.
\eeq
Here $|x_1,Q_1;\ldots;x_n,Q_n>$ means the configuration where a 
particle of class $Q_i (Q_i=1,2)$ 
is at position $x_i$ ($x_i =1,\ldots,N$). The summation $\{Q\} = \{Q_1,\ldots,
Q_n\}$ extends over all permutations of $n$ numbers 
in which $n_1$ terms are 1 and $n_2$ 
terms are 2, while the summation $\{x\} =\{x_1,\ldots,x_n\}$ runs, for each 
permutation $\{Q\}$, in the set of 
the $n$ nondecreasing 
integers satisfying 
\beq
\label{10}
x_{i+1} \geq x_{i} + s_{Q_i}, \quad i=1,\ldots,n-1, \quad s_{Q_1} 
\leq x_n -x_1 \leq N-s_{Q_n}.
\eeq
Before getting the results for general values of $n$ let us consider initially 
the cases where we have 1 or 2 particles.

{\it {\bf  n = 1.}}
For one particle on the chain (class 1 or 2),
as a consequence of the translational invariance 
of \rf{6} it is simple to verify directly that 
the eigenfunctions are the momentum-$k$ eigenfunctions 
\beq
\label{11}
|1,0> = \sum_{x=1}^{N} f(x,1) |x,1>, \quad \mbox {or} \quad 
|0,1> = \sum_{x=1}^{N} f(x,2) |x,2>, 
\eeq
with
\beq
\label{12}
 f(x,1) =f(x,2) =  e^{ikx},\quad k= \frac{2\pi l}{N}, 
l = 0,1,\ldots,N-1,
\eeq
and energy given by 
\beq
\label{13}
E = e(k) \equiv  -(\epsilon_-e^{ik} + \epsilon_+e^{-ik} -1).
\eeq

{\it {\bf  n =2.}} For two particles of classes $Q_1$ and 
$Q_2$ ($Q_1,Q_2=1,2$) on the lattice, the eigenvalue equation \rf{8} 
 gives  us two distinct relations depending on the relative location of 
the particles. The first relation applies to the case in which a 
particle of class $Q_1$ (size $s_{Q_1}$) is at position $x_1$ and a 
particle $Q_2$ (size $s_{Q_2}$) is at position $x_2$, where 
 $x_2 >x_1 +s_{Q_1}$. We obtain in this case the relation 
\bea
\label{14}
&&Ef(x_1,Q_1;x_2,Q_2) = -\epsilon_+f(x_1-1,Q_1;x_2,Q_2) - 
\epsilon_-f(x_1,Q_1;x_2+1,Q_2) \nonumber \\ 
&-&\epsilon_-f(x_1+1,Q_1;x_2,Q_2) 
 -\epsilon_+f(x_1,Q_1;x_2-1,Q_2) + 2f(x_1,Q_1;x_2,Q_2),
\eea
where we have used the relation $\epsilon_+ + \epsilon_- = 1$. This last 
equation
 can be solved promptly by the ansatz
\beq
\label{15}
f(x_1,Q_1;x_2,Q_2) = e^{ik_1x_1}e^{ik_2x_2},
\eeq
with energy
\beq
\label{16}
E = e(k_1) + e(k_2).
\eeq
Since this relation  is symmetric under the interchange of  $k_1$ and $k_2$, 
 we can write 
a more general solution of \rf{14} as
\bea
\label{17}
f(x_1,Q_1;x_2,Q_2) &=& \sum_P A_{P_1,P_2}^{Q_1,Q_2}e^{i(k_{P_1}x_1 + 
k_{P_2}x_2)}   \nonumber \\
&=& A_{1,2}^{Q_1,Q_2}e^{i(k_1x_1+k_2x_2)} + 
A_{2,1}^{Q_1,Q_2}e^{i(k_2x_1+k_1x_2)}
\eea
with the same energy as in \rf{16}. In \rf{17} the summation is over the  
permutations  $P=P_1,P_2$ of(1,2). The second relation applies when 
$x_2 = x_1 + s_{Q_1}$. In this case instead of \rf{14} we have 
\bea
\label{18}
Ef(x_1,Q_1&;&x_1 +s_{Q_1},Q_2) = -\epsilon_+f(x_1-1,Q_1;x_1+s_{Q_2},Q_2) - 
\epsilon_-f(x_1,Q_1;x_1+s_{Q_1}+1,Q_2)
 \nonumber \\
&- & \tilde \epsilon_{Q_1,Q_2}f(x_1,Q_2;x_1 +s_{Q_2},Q_1) + 
(1 + \tilde \epsilon_{Q_1,Q_2})f(x_1,Q_1;x_1 +s_{Q_1},Q_2),
\eea
where 
\beq
\label{19}
\tilde \epsilon_{1,1} =\tilde \epsilon_{2,2} = 0, \quad
\tilde \epsilon_{1,2} =\epsilon_+ \quad \mbox {and} \quad
\tilde \epsilon_{2,1} =\epsilon_-. 
\eeq
If we now substitute the ansatz \rf{17} with the energy \rf{16}, the 
constants $A_{12}^{Q_1,Q_2}$ and $A_{21}^{Q_1,Q_2}$, initially arbitrary, 
should now satisfy 
\beq
\label{20}
\sum_P \{ \left[ {\cal D}_{P_1,P_2} + e^{ik_{P_2}}(1- \tilde
\epsilon_{Q_1,Q_2})\right]e^{ik_{P_2}(s_{Q_1}-1)}A_{P_1,P_2}^{Q_1,Q_2} +  
\tilde \epsilon_{Q_1,Q_2} e^{ik_{P_2}s_{Q_2}}A_{P_1,P_2}^{Q_2,Q_1}\} = 0
\eeq
where

\beq
\label{21} 
{\cal D}_{i,j} = -(\epsilon_+ + \epsilon_-e^{i(k_i +k_j)}).
\eeq
At this point it is convenient to consider separately the case where $Q_1=Q_2$ 
from those where $Q_1 \neq Q_2$. If $Q_1=Q_2=Q$ ($Q=1,2$) eq. \rf{20} gives
\beq
\label{22}
\sum_P \left(  {\cal D}_{P_1,P_2} + e^{ik_{P_2}} 
\right)e^{ik_{P_2}(s_{Q}-1)}A_{P_1,P_2}^{Q_1,Q_2} = 0
\eeq
and the cases $Q_1 \neq Q_2$ give us the equations
\bea 
\label{23}
\sum_P\left[ \begin{array}{cc} {\cal D}_{P_1,P_2} +e^{ik_{P_2}}\epsilon_- &
\epsilon_+e^{ik_{P_2}}  \nonumber \\
\epsilon_-e^{ik_{P_2}} & {\cal D}_{P_1,P_2}+\epsilon_+e^{ik_{P_2}} \nonumber 
\end{array} \right]
\left[ \begin{array}{c} e^{ik_{P_2}(s_1-1)}A_{P_1,P_2}^{1,2} \nonumber \\
e^{ik_{P_2}(s_2-1)}A_{P_1,P_2}^{2,1} \end{array} \right] = 0.
\eea  
Performing the above summation we obtain, after lengthy but straightforward 
algebra, the following relation among the amplitudes 
\bea  \label{24}
\left[ \begin{array}{c} A_{1,2}^{1,2}e^{ik_2(s_1-1)} \nonumber \\
A_{1,2}^{2,1}e^{ik_2(s_2-1)} \end{array} \right] = 
-\frac{{\cal D}_{1,2} + e^{ik_1}} {{\cal D}_{1,2} +e^{ik_2}} 
\left\{ 1 - \Phi(k_1,k_2)\left[ \begin{array} {cc} \epsilon_+ & -\epsilon_+
\nonumber \\
 -\epsilon_- & \epsilon_- \end{array} \right] \right\}
\left[ \begin{array} {c} 
A_{2,1}^{12}e^{ik_1(s_1-1)} \nonumber \\
A_{2,1}^{21}e^{ik_1(s_2-1)} \nonumber \end{array} \right],
\eea
where
\beq \label{25}
\Phi(k_1,k_2) = \frac{e^{ik_1} - e^{ik_2}}{{\cal D}_{1,2} +e^{ik_1}} .
\eeq
Equations \rf{22} and \rf{24} can be written in a compact form 
\beq \label{26}
A_{P_1,P_2}^{Q1,Q2} = -\Xi_{P_1,P_2} \sum_{Q_1',Q_2' =1}^2 
S_{Q_1',Q_2'}^{Q_1,Q_2}(k_{P_1},k_{P_2})A_{P_2,P_1}^{Q_2',Q_1'}, \quad 
(Q_1,Q_2 =1,2)
\eeq
with
\beq \label{27}
\Xi_{l,j} = \frac{{\cal D}_{l,j} + e^{ik_l}} {{\cal D}_{l,j} +e^{ik_j}} =
\frac {\epsilon_+ + \epsilon_-e^{i(k_l+k_j)} -e^{ik_l}} 
{\epsilon_+ + \epsilon_-e^{i(k_l+k_j)} - e^{ik_j}},
\eeq
where we have introduced the $S$ matrix. From \ref{22} and \rf{24}, the only 
non zero elements of $S_{Q_1',Q_2'}^{Q_1,Q_2}$ are:
\bea \label{28}
S_{1,1}^{1,1}(k_1,k_2) 
&=& e^{i(k_1-k_2)(s_1-1)}, \quad s_{2,2}^{2,2}(k_1,k_2)= 
e^{i(k_1-k_2)(s_2-1)}, \nonumber \\
S_{2,1}^{1,2}(k_1,k_2)
 &=& \left[ 1-\epsilon_+\Phi(k_1,k_2)\right] e^{i(k_1-k_2)(s_1-1)},
\nonumber \\
S_{1,2}^{2,1}(k_1,k_2) &=& 
\left[ 1-\epsilon_-\Phi(k_1,k_2)\right] e^{i(k_1-k_2)(s_2-1)},
\nonumber \\
S_{1,2}^{1,2} (k_1,k_2)&=& 
\epsilon_+\Phi(k_1,k_2) e^{ik_1(s_2-1)}e^{-ik_2(s_1-1)},
\nonumber \\
S_{2,1}^{2,1} (k_1,k_2)&=& 
\epsilon_-\Phi(k_1,k_2) e^{ik_1(s_1-1)}e^{-ik_2(s_2-1)}.
\eea
A graphical representation of the S matrix is shown in Fig. 2a. Equations 
\rf{26} do not fix the ``wave numbers" $k_1$ and $k_2$. In general, these numbers 
are complex, and are fixed due to the cyclic boundary condition 
\beq
\label{29}
f(x_1,Q_1;x_2,Q_2) =  f(x_2,Q_2;x_1+N,Q_1),
\eeq
which from \rf{17} gives the relation
\beq
\label{30}
A_{1,2}^{Q_1Q_2}  = e^{ik_1N}A_{2,1}^{Q_2,Q_1}, \quad \quad A_{2,1}^{Q_1,Q_2} = 
e^{ik_2N}A_{2,1}^{Q_2,Q_1}.
\eeq
This last equation, when solved by exploiting \rf{26}-\rf{28}, gives us the 
possible values of $k_1$ and $k_2$, and from \rf{16} the eigenenergies 
in the sector with 2 particles. Instead of solving these equations for the 
particular case $n=2$ let us now consider the case of general $n$.

{\it {\bf General n.}} The above calculation can be generalized for arbitrary 
values of $n_1$ and $n_2$ of particles of classes 1 and 2, respectively 
($n_1 +n_2 = n$).  The ansatz for the wave function \rf{9}  becomes 
\beq
\label{31}
f(x_1,Q_1;\ldots;x_n,Q_n) = \sum_P A_{P_1,\ldots,P_n}^{Q_1,\cdots, Q_n}
 e^{i(k_{P_1}x_1+\cdots + 
k_{P_n}x_n)},
\eeq
where the sum extends over all permutations $P$ of the integers 
$1,2,\ldots,n$. For the components $|x_1,Q_1;\ldots;x_n,Q_n>$ where  
$x_{i+1} -x_i >s_{Q_i}$ for $i=1,2,\ldots,n$, it is simple to see
that the eigenvalue equation \rf{8} is satisfied by the ansatz \rf{31} 
with energy 
\beq
\label{32}
E = \sum_{j=1}^n e(k_j).
\eeq
On the other hand if
 a pair of particles of class $Q_i,Q_{i+1}$ is at positions 
$x_i,\; x_{i+1}$, where $x_{i+1} = 
x_i +s_{Q_i}$, equation \rf{8} with the ansatz \rf{31} and the relation 
\rf{32} give us the generalization of relation \rf{26}, namely
\beq
\label{33}
A_{\ldots,P_i,P_{i+1},\ldots}^{\cdots, Q_i,Q_{i+1},\cdots} = 
 -\Xi_{P_i,P_{i+1}} \sum_{Q_1',Q_2'}^2 S_{Q_1',Q_2'}^{Q_i,Q_{i+1}}
(k_{P_i},k_{P_{i+1}}) A_{\ldots, P_{i+1},P_i,\ldots}^{\cdots, Q_2',Q_1',\cdots} 
\quad \;\; Q_i,Q_{i+1} =1,2,
\eeq
with $S$ given by eq. \rf{28}. Inserting the ansatz \rf{31} in the 
boundary condition 
\beq
\label{34}
f(x_1,Q_1;\ldots;x_n,Q_n) = f(x_2,Q_2;\ldots;x_n,Q_n;x_1+N,Q_1)
\eeq
we obtain  the additional relation 
\beq
\label{35}
A_{P_1,\ldots,P_n}^{Q_1,\cdots, Q_n} = e^{ik_{P_1}N}
A_{P_2,\ldots,P_n,P_1}^{Q_2,\cdots, Q_n,Q_1},
\eeq
which together with \rf{33} should give us the energies.

 Successive applications of \rf{33} give us in 
general distinct relations between the amplitudes. For example $A_{\ldots,
k_1,k_2,k_3,\ldots}^{\ldots,\alpha,\beta,\gamma,\ldots}$ relate to 
$A_{\ldots,k_3,k_2,k_1,\ldots}^{\ldots,\gamma,\beta,\alpha,\ldots}$ by 
performing the permutations $\alpha\beta\gamma \rightarrow \beta\alpha\gamma 
\rightarrow \beta\gamma\alpha \rightarrow \gamma\beta\alpha$ or 
 $\alpha\beta\gamma \rightarrow \alpha\gamma\beta  
\rightarrow \gamma\alpha\beta \rightarrow \gamma\beta\alpha$, and consequently 
the $S$-matrix should satisfy the Yang-Baxter \cite{yang,baxter} equation
\bea
\label{36}
\sum_{\gamma,\gamma',\gamma''=1}^2 S_{\gamma,\gamma'}^{\alpha,\alpha'}(k_1,k_2)
S_{\beta,\gamma''}^{\gamma,\alpha''}(k_1,k_3)& &S_{\beta',\beta''}^
{\gamma',\gamma''}(k_2,k_3) = \nonumber \\
& & \sum_{\gamma,\gamma',\gamma''=1}^2
 S_{\gamma',\gamma''}^{\alpha',\alpha''}(k_2,k_3)
S_{\gamma,\beta''}^{\alpha,\gamma''}(k_1,k_3)S_{\beta,\beta'}^
{\gamma,\gamma'}(k_1,k_2), 
\eea
for $\alpha,\alpha',\alpha'',\beta,\beta',\beta'' =1,2$ and 
 $S$ given by \rf{28}. In Fig. 2b we 
 show graphically this equation. Actually the relation \rf{36} 
is a necessary and sufficient condition \cite{yang,baxter} 
to obtain a non-trivial 
solution for the amplitudes in Eq. \rf{33}. 

We can verify by a long and straightforward calculation that 
for arbitrary values of $s_1$ and $s_2$,
the $S$ matrix \rf{28}, satisfies the Yang-Baxter 
equation \rf{36}, and consequently we may use relations \rf{33} and \rf{35} 
to obtain the eigenenergies of the Hamiltonian \rf{6}. Applying relation  
\rf{33} $n$ times on the right of equation \rf{35} we obtain a relation 
between the amplitudes with the same momenta. Using the graphical 
representation in Fig. 3a for the $S$ matrix, we illustrate in Fig. 4 the 
result of such applications. We then obtain
\bea \label{37}
A_{P_1,\ldots,P_n}^{Q_1,\ldots,Q_n} = e^{ik_{P_1}N} A_{P_2,\ldots,P_n,P_1}^
{Q_2,\ldots,Q_n,Q_1} = \left(\prod_{i=2}^n -\Xi_{P_i,P_1}\right) 
e^{ik_{P_1}N} \sum_{Q_1',\ldots,Q_n'}\sum_{Q_1'',\ldots,Q_n''} \nonumber \\
S_{Q_1',Q_1''}^{Q_1,Q_2''}(k_{P_1},k_{P_1})
S_{Q_2',Q_2''}^{Q_2,Q_3''}(k_{P_2},k_{P_1})\cdots
S_{Q_{n-1}',Q_{n-1}''}^{Q_{n-1},Q_n''}(k_{P_{n-1}},k_{P_1})
S_{Q_{n}',Q_{n}''}^{Q_{n},Q_1''}(k_{P_{n}},k_{P_1})
A_{P_1,\ldots,P_n}^{Q_1',\ldots,Q_n'},
\eea
where we have introduced the harmless extra sum
\beq
\label{38}
1 = \sum_{Q_1'',Q_2''=1}^2 \delta_{Q_2'',Q_1'} \delta_{Q_1'',Q_1} =
 \sum_{Q_1'',Q_2''=1}^2 S_{Q_1',Q_1''}^{Q_1,Q_2''}(k_{P_1},k_{P_1}).
\eeq
In order to fix the values of $\{k_j\}$ we should solve \rf{37}, i.e., 
we should find the eigenvalues $\Lambda(k)$ of the matrix
\beq
\label{39}
T(k)_{\{Q'\}}^{\{Q\}} = \sum_{Q_1'',\ldots,Q_n''=1}^2 \left\{ \left(
\prod_{l=1}^{n-1} 
S_{Q_l',Q_l''}^{Q_l,Q_{l+1}''}(k_{P_l} ,k)\right) 
S_{Q_n',Q_n''}^{Q_n,Q_1''}(k_{P_n},k) \right\},
\eeq
and the Bethe-ansatz equations which fix the set $\{k_l\}$ will be given from
 \rf{37} by 
\beq \label{40}
e^{-ik_jN} = (-1)^{n-1}\left( 
\prod_{l=1}^n\Xi_{l,j}\right)\Lambda(k_j), \quad j=1,\ldots,n.
\eeq
Using the graphical representation for $S$ given 
in Fig.~3a the  matrix $T$ in \rf{39} can be represented graphically  
as in Fig.~5, 
and we identify $T(k)$ as the transfer matrix of an inhomogeneous 
6-vertex model, on a periodic lattice,
with Boltzmann weights  $S_{Q_1',Q_2'}^{Q_1,Q_2}(k_{P_l},k)$ ($l=1,\ldots,n$).  
Consequently, in order to obtain the eigenenergies of the quantum 
chain \rf{6} we should diagonalize the above transfer matrix $T(k)$.

\noindent {\bf Diagonalization of $T(k)$} \\
The simplest way to diagonalize $T$ is through the introduction of the 
monodromy matrix ${\cal M} (k)$ \cite{takh}, which is a transfer matrix of the 
inhomogeneous vertex model under consideration, where, instead of 
being periodic, the first and last link in the horizontal direction are 
fixed to the values $\mu_1$ and $\mu_{n+1}$ ($\mu_1, \mu_{n+1} =1,2$), that is 
\beq \label{41}
{\cal M}_{\{Q'\},\mu_1}^{\{Q\},\mu_{n+1}}(k) = \sum_{\mu_2,\ldots,\mu_n} 
S_{Q_1',\mu_1}^{Q_1,\mu_2}(k_{P_1},k)
S_{Q_2',\mu_2}^{Q_2,\mu_3}(k_{P_2},k) \cdots 
S_{Q_{n-1}',\mu_{n-1}}^{Q_{n-1},\mu_n}(k_{P_{n-1}},k)
S_{Q_{n}',\mu_{n}}^{Q_{n},\mu_{n+1}}(k_{P_{n}},k).
\eeq
The monodromy matrix ${\cal M}_{\{Q'\},\mu_1}^{\{Q\},\mu_{n+1}}(k)$ is 
represented in Fig. 6a and has coordinates $\{Q\},\{Q'\}$ in the vertical 
space ($2^n$ dimensions) and coordinates $\mu_1,\mu_{n+1}$ in the 
horizontal space (4 dimensions).  
This matrix satisfies the following important relations 
\beq \label{42}
S_{\nu_1,\mu_1}^{\nu_1',\mu_1'}(k',k)
{\cal M}_{\{\alpha_l\},\mu_1'}^{\{\gamma_l\},\mu_{n+1}}(k)
{\cal M}_{\{\gamma_l\},\nu_1'}^{\{\beta_l\},\nu_{n+1}}(k') = 
{\cal M}_{\{\alpha_l\},\nu_1}^{\{\gamma_l\},\nu_{n+1}'}(k')
{\cal M}_{\{\gamma_l\},\mu_1}^{\{\beta_l\},\mu_{n+1}'}(k)  
S_{\nu_{n+1}',\mu_{n+1}'}^{\nu_{n+1},\mu_{n+1}}(k',k).
\eeq
This relation is shown graphically in Fig. 6b; its validity follows 
directly from successive applications of the Yang-Baxter equations \rf{36} 
(see Fig. 2b). 

In order to exploit relation \rf{42} let us denote the components of the  
monodromy matrix in the horizontal space by
\bea \label{43}
A(k)& =& A(k)_{\{\alpha_l\}}^{\{\gamma_l\}} = {\cal M}_{\{\alpha_l\},1}^
{\{\gamma_l\},1}(k), \quad 
B(k) = B(k)_{\{\alpha_l\}}^{\{\gamma_l\}} = {\cal M}_{\{\alpha_l\},1}^
{\{\gamma_l\},2}(k), \quad \nonumber \\
C(k)& =& C(k)_{\{\alpha_l\}}^{\{\gamma_l\}} = {\cal M}_{\{\alpha_l\},2}^
{\{\gamma_l\},1}(k), \quad 
D(k) = D(k)_{\{\alpha_l\}}^{\{\gamma_l\}} = {\cal M}_{\{\alpha_l\},2}^
{\{\gamma_l\},2}(k),
\eea
and clearly the transfer matrix $T(k)$ of the periodic inhomogeneous 
lattice, we want to diagonalize, is given by 
\beq \label{44}
T(k) = A(k) + D(k).
\eeq
As a consequence of \rf{42} the matrices $A$, $B$, $C$ and $D$ in \rf{43} 
obey some algebraic relations. The elements $(\nu_1,\mu_1,\nu_{n+1},\mu_{n+1}) 
= (1,1,2,2)$, $(1,1,1,2)$ and 
$(1,2,2,2)$ give us the relations 
\beq \label{42a}
[B(k),B(k')] =0,
\eeq
\beq \label{43a}
A(k)B(k') = \frac{S_{1,1}^{1,1}(k,k')}{S_{1,2}^{1,2}(k,k')}B(k')A(k) -
 \frac{S_{2,1}^{1,2}(k,k')}{S_{1,2}^{1,2}(k,k')}B(k)A(k'), 
\eeq
\beq \label{44a}
D(k)B(k') = \frac{S_{2,2}^{2,2}(k',k)}{S_{1,2}^{1,2}(k',k)}B(k')D(k) -
 \frac{S_{1,2}^{2,1}(k',k)}{S_{1,2}^{1,2}(k',k)}B(k)D(k'), 
\eeq
respectively. The diagonalization of $T(k)$ in \rf{44} will be done by
exploiting the above relations. This procedure is known in the literature 
as the algebraic Bethe ansatz \cite{takh}. The first step in this method follows 
from the identification of a reference state $|\Omega>$, which should be  
an eigenstate of $A(k)$ and $D(k)$, and hence $T(k)$, but not of $B(k)$.  
In our problem it is simple to see that a possible reference state is 
$|\Omega> = |\{\alpha_l=1\}>_{l=1,\ldots,n}$, which corresponds to a state 
with first-class particles only. It is simple to calculate 
\bea \label{46}
A(k)|\Omega> &=& a(k)|\Omega >, \quad \quad 
D(k)|\Omega> =d(k)|\Omega >, \nonumber \\
C(k)|\Omega> &=& 0, \quad \quad 
B(k)|\Omega> =\sum_{i=1}^nb_i(k)|\Omega >, 
\eea
where 
\bea \label{47}
a(k) &=& \prod_{i=1}^nS_{1,1}^{1,1}(k_{P_i},k), \quad \quad 
d(k) = \prod_{i=1}^nS_{2,1}^{1,2}(k_{P_i},k), \quad \quad \nonumber \\
b_i(k) &=& \prod_{l=1}^{i-1}S_{1,1}^{1,1}(k_{P_l},k)
\prod_{l=j}^nS_{1,2}^{1,2}(k_{P_l},k).  
\eea
In Fig. 7 we show a graphical representation of the above relation. The idea 
in the algebraic Bethe ansatz is that $B(k)$ acts as a creation operator  
in the reference (``vacuum") state, i.e.,  it creates particles of 
second class in a sea of particles of first class $|\Omega>$, and it is 
expected that a general eigenvector of $T(k)$ in the sector with $n_2$ 
second-class particles will be given by the ansatz
\beq \label{48}
\Psi(k_1^{(1)},k_2^{(1)},\ldots,k_{n_2}^{(1)}) = B(k_1^{(1)})B(k_2^{(1)})\cdots 
B(k_{n_2}^{(1)})|\Omega >.
\eeq
The numbers $\{k_i^{(1)}\}, i=1,\ldots,n_2$ here play the role of the 
``wave number" $\{k_i\}$ in the coordinate Bethe ansatz \rf{31}, and are going 
to be fixed by the eigenvalue equation
\beq \label{49}
T(k)\Psi(k_1^{(1)},\ldots,k_{n_2}^{(1)}) = 
\left( A(k) + D(k) \right) \Psi(k_1^{(1)},\ldots,k_{n_2}^{(1)}) = 
\Lambda(k)\Psi(k_1^{(1)},\ldots,k_{n_2}^{(1)}).
\eeq
Before deriving the relation for general values of $n_2$, let us consider 
initially the cases where $n_2=1$ and $n_2 =2$.

\noindent {\bf $\bf{n_2 =1}$}. Using relation \rf{43a} and \rf{44a} in \rf{49} we 
obtain 
\bea \label{50}
T(k)\Psi(k_1^{(1)}) &=& 
\left[ \frac{S_{1,1}^{1,1}(k,k_1^{(1)})}{S_{1,2}^{1,2}(k,k_1^{(1)})} +
d(k)\frac{S_{2,2}^{2,2}(k_1^{(1)},k)}{S_{1,2}^{1,2}(k_1^{(1)},k)} \right] 
\Psi(k_1^{(1)}) \nonumber \\
&-&\left[\frac{S_{2,1}^{1,2}(k,k_1^{(1)})}{S_{1,2}^{1,2}(k,k_1^{(1)})}  
+ d(k_1^{(1)})\frac{S_{1,2}^{2,1}(k_1^{(1)},k)}{S_{1,2}^{1,2}(k_1^{(1)},k)} 
\right] B(k) |\Omega> = \Lambda(k)\Psi(k_1^{(1)}),
\eea
which is clearly satisfied if the coefficient of $B(k)|\Omega >$ (unwanted 
term) vanishes. This condition, which fixes $k_1^{(1)}$, is given by 
\beq \label{51}
\prod_{j=1}^n S_{1,2}^{1,2}(k_{P_j},k_1^{(1)}) =1,
\eeq
where we used the relation 
\beq \label{52}
\frac{S_{2,1}^{1,2}(k,k')} {S_{1,2}^{1,2}(k,k')} = 
- \frac{S_{1,2}^{2,1}(k',k)} {S_{1,2}^{1,2}(k',k)} ,
\eeq
valid for the $S$ matrix \rf{28}. The eigenvalue is given by 
\beq \label{53}
\Lambda(k) = 
\frac{S_{1,1}^{1,1}(k,k_1^{(1)})}{S_{1,2}^{1,2}(k,k_1^{(1)})} 
+ \frac{S_{2,2}^{2,2}(k_1^{(1)},k)}{S_{1,2}^{1,2}(k_1^{(1)},k) }
\prod_{j=1}^nS_{1,2}^{1,2}(k_{P_j},k),
\eeq
provided \rf{51}is satisfied.

\noindent  $\bf{n_2 =2}$. 
In this case the application of $A(k)$ and $D(k)$ in 
the ansatz \rf{48} give us
\bea \label{54} 
A(k)\Psi(k_1^{(1)},k_2^{(1)}) &=&
\frac{S_{1,1}^{1,1}(k,k_1^{(1)})S_{1,1}^{1,1}(k,k_2^{(1)})}
{S_{1,2}^{1,2}(k,k_1^{(1)})S_{1,2}^{1,2}(k,k_2^{(1)})}
a(k)\Psi(k_1^{(1)},k_2^{(1)}) 
\nonumber \\
&-& \frac{S_{2,1}^{1,2}(k,k_1^{(1)})S_{1,1}^{1,1}(k_1^{(1)},k_2^{(1)})}
{S_{1,2}^{1,2}(k,k_1^{(1)})S_{1,2}^{1,2}(k_1^{(1)},k_2^{(1)})}
a(k_1^{(1)})B(k)B(k_2^{(1)})|\Omega > \nonumber \\
&-& \frac{S_{2,1}^{1,2}(k,k_2^{(1)})S_{1,1}^{1,1}(k_2^{(1)},k_1^{(1)})}
{S_{1,2}^{1,2}(k,k_2^{(1)})S_{1,2}^{1,2}(k_2^{(1)},k_1^{(1)})}
a(k_2^{(1)})B(k)B(k_1^{(1)})|\Omega >, 
\eea
and 
\bea \label{55} 
D(k)\Psi(k_1^{(1)},k_2^{(1)}) &=&
\frac{S_{2,2}^{2,2}(k_1^{(1)},k)S_{2,2}^{2,2}(k_2^{(1)},k)}
{S_{1,2}^{1,2}(k_1^{(1)},k)S_{1,2}^{1,2}(k_2^{(1)},k)}
d(k)\Psi(k_1^{(1)},k_2^{(1)}) \nonumber \\
&-& \frac{S_{1,2}^{2,1}(k_1^{(1)},k)S_{2,2}^{2,2}(k_2^{(1)},k_1^{(1)})}
{S_{1,2}^{1,2}(k_1^{(1)},k)S_{1,2}^{1,2}(k_2^{(1)},k_1^{(1)})}
d(k_1^{(1)})B(k)B(k_2^{(1)})|\Omega > \nonumber \\
&-& \frac{S_{1,2}^{2,1}(k_2^{(1)},k)S_{2,2}^{2,2}(k_1^{(1)},k_2^{(1)})}
{S_{1,2}^{1,2}(k_2^{(1)},k)S_{1,2}^{1,2}(k_1^{(1)},k_2^{(1)})}
d(k_2^{(1)})B(k)B(k_1^{(1)})|\Omega >, 
\eea
where besides the relation \rf{43a} and \rf{44a} we have used \rf{42a}, which 
imply $\Psi(k_1^{(1)},k_2^{(1)}) = \Psi(k_2^{(1)},k_1^{(1)})$. From \rf{54} 
and \rf{55} the condition \rf{49} gives us
\beq \label{56}
\Lambda(k) = 
\prod_{i=1}^n S_{1,1}^{1,1} (k_{P_i},k) 
\prod_{j=1}^2 \frac{S_{1,1}^{1,1}(k,k_j^{(1)})} {S_{1,2}^{1,2}(k,k_j^{(1)})} 
+ \prod_{i=1}^n S_{1,2}^{1,2} (k_{P_i},k) 
\prod_{j=1}^2 \frac{S_{2,2}^{2,2}(k_j^{(1)},k)} {S_{1,2}^{1,2}(k_j^{(1)},k)}, 
\eeq
under the condition, which fixes $k_1^{(1)}$ and $k_2^{(1)}$, 
\beq \label{57}
\prod_{i=1}^n S_{1,2}^{1,2}(k_{P_i},k_{\alpha}^{(1)}) = 
\prod_{\beta =1 (\neq \alpha)}^2 
\frac{S_{1,1}^{1,1}(k_{\alpha}^{(1)},k_{\beta}^{(1)})
S_{1,2}^{1,2}(k_{\beta}^{(1)},k_{\alpha}^{(1)})}
{S_{1,2}^{1,2}(k_{\alpha}^{(1)},k_{\beta}^{(1)})
S_{2,2}^{2,2}(k_{\beta}^{(1)},k_{\alpha}^{(1)})}
\prod_{j=1}^nS_{1,1}^{1,1}(k_{P_j},k_{\alpha}^{(1)}) \quad 
\alpha =1,2.
\eeq
In deriving this last expression the relation \rf{52}
 was also used.

\noindent {\bf General $\bf{n_2}$.} The previous procedure can be iterated 
straightforwadly for arbitrary numbers $n_2$ of second-class particles, 
which gives
\bea \label{58}
A(k)\Psi(k_1^{(1)},\ldots,k_{n_2}^{(1)}) & = & \Lambda^{(A)}(k)
\Psi(k_1^{(1)},\ldots,k_{n_2}^{(1)}) \nonumber \\  
&-& \sum_{i=1}^n\Lambda_i^{(A)}(k) \Psi(k,k_1^{(1)},\ldots,k_{i-1}^{(1)},
k_{i+1}^{(1)},\ldots,k_{n_2}^{(1)}), \nonumber \\ 
D(k)\Psi(k_1^{(1)},\ldots,k_{n_2}^{(1)}) & = & \Lambda^{(D)}(k)
\Psi(k_1^{(1)},\ldots,k_{n_2}^{(1)})  \nonumber \\
&-& \sum_{i=1}^n\Lambda_i^{(D)}(k) \Psi(k,k_1^{(1)},\ldots,k_{i-1}^{(1)},
k_{i+1}^{(1)},\ldots, k_{n_2}^{(1)}), 
\eea
where
\bea \label{59}
\Lambda^{(A)}(k) &=& \prod_{j=1}^nS_{1,1}^{1,1}(k_{P_j},k) 
\prod_{i=1}^{n_2} \frac 
{S_{1,1}^{1,1}(k,k_i^{(1)})} {S_{1,2}^{1,2}(k,k_i^{(1)})}, \nonumber \\
\Lambda^{(D)}(k) &=& \prod_{j=1}^nS_{1,2}^{1,2}(k_{P_j},k) 
\prod_{i=1}^{n_2} \frac 
{S_{2,2}^{2,2}(k_i^{(1)},k)} {S_{1,2}^{1,2}(k_i^{(1)},k)}, \nonumber \\
\Lambda_j^{(A)}(k) &=& \frac{S_{2,1}^{1,2}(k,k_j^{(1)})}
{S_{1,2}^{1,2}(k,k_j^{(1)})}
a(k_j^{(1)})  \prod_{l=1(\neq j)}^{n_2}
\frac {S_{1,1}^{1,1}(k_j^{(1)},k_l^{(1)})}
 {S_{1,2}^{1,2}(k_j^{(1)},k_l^{(1)})}, \nonumber \\
\Lambda_j^{(D)}(k) &=& \frac{S_{1,2}^{2,1}(k_j^{(1)},k)}
{S_{1,2}^{1,2}(k_j^{(1)},k)}
d(k_j^{(1)})  \prod_{l=1(\neq j)}^{n_2}
\frac {S_{2,2}^{2,2}(k_l^{(1)},k_j^{(1)})}
 {S_{1,2}^{1,2}(k_l^{(1)},k_j^{(1)})}. \nonumber \\
\eea
Then $\Psi(k_1^{(1)},\ldots,k_{n_2}^{(1)})$ is an eigenvector of $T$ with 
eigenvalue $\Lambda(k) = \Lambda^{(A)}(k) + \Lambda^{(D)}(k)$, i. e., 
\beq \label{60}
\Lambda(k) = \prod_{i=1}^nS_{1,1}^{1,1}(k_{P_i},k) 
\prod_{\alpha=1}^{n_2} \frac 
{S_{1,1}^{1,1}(k,k_{\alpha}^{(1)})} {S_{1,2}^{1,2}(k,k_{\alpha}^{(1)})} + 
 \prod_{i=1}^nS_{1,2}^{1,2}(k_{P_i},k] 
\prod_{\alpha=1}^{n_2} \frac 
{S_{2,2}^{2,2}(k_{\alpha}^{(1)},k)} {S_{1,2}^{1,2}(k_{\alpha}^{(1)},k)},
\eeq
if the following conditions, which fix $\{k_1^{(1)},\ldots,k_{n_2}\}$ are  
satisfied:
\beq \label{61a}
\Lambda_i^{(A)}(k) +\Lambda_i^{(D)}(k) = 0, \quad i=1,\ldots,n_2.
\eeq
Using the relations \rf{52}, this last condition can be written as 
\beq \label{61}
\prod_{j=1}^n \frac 
{S_{1,2}^{1,2}(k_{P_j},k_{\alpha}^{(1)})} 
{S_{1,1}^{1,1}(k_{P_j},k_{\alpha}^{(1)})} = 
\prod_{\beta =1 (\neq \alpha)}^{n_2} 
\frac{S_{1,1}^{1,1}(k_{\alpha}^{(1)},k_{\beta}^{(1)})
S_{1,2}^{1,2}(k_{\beta}^{(1)},k_{\alpha}^{(1)})}
{S_{1,2}^{1,2}(k_{\alpha}^{(1)},k_{\beta}^{(1)})
S_{2,2}^{2,2}(k_{\beta}^{(1)},k_{\alpha}^{(1)})}, \quad 
\alpha =1,\ldots,n_2,
\eeq
which concludes the diagonalization of $T(k)$.

Now let us return to our original problem of finding the eigenvalues 
of the Hamiltonian \rf{6}. The Bethe-ansatz equations will be obtained 
by inserting in \rf{40} the eigenvalues evaluated at $k_j$, i. e., 
$\Lambda(k_j)$, given in \rf{60}, with the condition \rf{61}. 
Taking into account that $S_{1,2}^{1,2}(k,k) =0$, we obtain 
\beq \label{62}
e^{-ik_jN} = (-)^{n-1}\prod_{l=1}^n \left(\Xi_{l,j} 
S_{1,1}^{1,1}(k_l,k_j)\right) \prod_{\alpha=1}^{n_2} 
\frac{S_{1,1}^{1,1}(k_j,k_{\alpha}^{(1)})}
{S_{1,2}^{1,2}(k_j,k_{\alpha}^{(1)})}, \quad j=1,\ldots,n,
\eeq
with the condition 
\beq \label{63}
\prod_{j=1}^n \frac 
{S_{1,2}^{1,2}(k_{j},k_{\alpha}^{(1)})} 
{S_{1,1}^{1,1}(k_{j},k_{\alpha}^{(1)})} = 
\prod_{\beta =1 (\neq \alpha)}^{n_2} 
\frac{S_{1,1}^{1,1}(k_{\alpha}^{(1)},k_{\beta}^{(1)})
S_{1,2}^{1,2}(k_{\beta}^{(1)},k_{\alpha}^{(1)})}
{S_{1,2}^{1,2}(k_{\alpha}^{(1)},k_{\beta}^{(1)})
S_{2,2}^{2,2}(k_{\beta}^{(1)},k_{\alpha}^{(1)})}, \quad 
\alpha =1,\ldots,n_2.
\eeq
Writing explicitly the $S$-matrix elements \rf{28} in \rf{62} and \rf{63} 
we can state that the energies, in the sector with $n_1$ first-class particles
and $n_2$ second-class particles are given by
\beq \label{64}
E = -\sum_{j=1}^n(\epsilon_-e^{ik_j} +\epsilon_+e^{-ik_j} -1),
\eeq
where $\{k_j, j =1,\ldots,n\}$ are given by the solutions of 
\bea \label{65}
e^{ik_j(N-n_2(s_2-s_1))} &=& (-)^{n_1-1}  \prod_{l=1}^n 
\left[ \left( \frac{e^{ik_j}}{e^{ik_l}} \right)^{s_1-1} 
\frac{\epsilon_+ + \epsilon_-e^{i(k_l+k_j)} -e^{ik_j}}
{\epsilon_+ + \epsilon_-e^{i(k_l+k_j)} -e^{ik_l}}\right] \nonumber \\
&\times& \prod_{\alpha=1}^{n_2} \frac{\epsilon_+ (e^{ik_j} -
e^{ik_{\alpha}^{(1)}}  )}
{\epsilon_+ + \epsilon_-e^{i(k_j +k_{\alpha}^{(1)})} -e^{ik_{\alpha}^{(1)}}}
\quad \quad (j=1,\ldots,n),
\eea
\bea \label{66}
e^{ik_j(s_2-s_1)n} &(\epsilon_+)^n&\prod_{i=1}^n \frac {e^{ik_j}-
e^{ik_{\alpha}^{(1)}}}
{\epsilon_+ + \epsilon_-e^{i(k_j+k_{\alpha}^{(1)})} -e^{ik_j}} = 
 \nonumber \\ 
&(-)^{n_1-1}&\prod_{\beta=1}^{n_2} 
\frac{\epsilon_+ + \epsilon_-e^{i(k_{\alpha}^{(1)} +k_{\beta}^{(1)})} 
-e^{ik_{\alpha}^{(1)}}}
{\epsilon_+ + \epsilon_-e^{i(k_{\alpha}^{(1)} +k_{\beta}^{(1)})} 
-e^{ik_{\beta}^{(1)}}} \quad \quad (\alpha=1,\ldots,n_2).
\eea
It is interesting to observe that in the particular case where $n_2=0$ we 
obtain the Bethe-ansatz equations, recently derived \cite{alc-bar1}, for the 
asymmetric diffusion problem with particles of size $s_1$. Also the case  
$s_1=s_2=1$ give us the corresponding Bethe ansatz equations for the  
standard problem of second class particles. The Bethe-ansatz equations for 
the fully asymmetric problem is obtained by setting $\epsilon_+=1$ and
 $\epsilon_- =0$.

\section{ Conclusions and generalizations }

We obtained through the Bethe ansatz the exact solution of the problem of 
first- and second-class particles diffusing and interchanging positions on 
the lattice. We show that the solution can be derived in the general case 
where the particles have arbitrary sizes. 

Some extensions of our results can be made. A first  generalization is the 
problem where $M>2$ hierarchical ordered classes of particles 
(1st, 2nd, ...,Mth classes), with sizes $s_i$ ($i=1,\ldots,M$),  besides 
diffusing on the lattice, interchange positions in such way that particles 
of higher hierarchical classes see the lower ones in the same way as they 
see the holes (0 class). The allowed processes and the Hamiltonians are 
simple generalizations of relations \rf{2}-\rf{4} and \rf{6}, respectively. 
The Hamiltonian obtained in a open lattice will be $U_q(SU(M))$ symmetric 
and in the case $s_1 =s_2 =\ldots=s_M=1$ it is related with the spin 
$S=\frac{M}{2}$ Sutherland model \cite{sutherland}. The model is certainly integrable and 
its eigenspectra will be given in terms of $M$-nested Bethe ansatz equations 
which are generalizations of \rf{65}-\rf{66}.

A further quite interesting generalization of our model happens when we 
consider molecules in one or both classes with size $s=0$. Molecules 
of size zero do not occupy space on the lattice, having no hard-core 
exclusion effect. Consequently we may have, at a given lattice point,  
an arbitrary number of them. The Bethe-ansatz solution presented in the previous 
section is extended directly in this case (the equations are the same) 
and the eigenenergies are given 
by fixing in \rf{64}-\rf{66} the appropriate sizes of the molecules. It 
is interesting to remark that particles of second class with size $s_2=0$, 
contrary to the case  $s_2>1$, where they ``accelerate" the diffusion of the 
first class particles, now they ``retard" the diffusive motion of these 
particles. The quantum Hamiltonian in the cases where the particles have  
size zero are written in terms of spin $s=\infty$ quantum chains.

The Bethe-ansatz equations in the case of the asymmetric 
diffusion, with particles of unit size \cite{spohn,kim}, or with 
arbitrary size \cite{alc-bar1}, 
was used to obtain the finite-size corrections of the 
mass gap $G_N$ of the associated quantum chain. The real part of these 
finite-size corrections are governed by the dynamical critical exponent 
$z$, i. e., 
\beq
\mbox {Re}(G_N) \sim N^{-z}. \nonumber
\eeq
The calculation of the exponent $z$ for the model presented in this paper, 
with particles of arbitrary sizes, is presently in progress \cite{alc-bar2}.

\acknowledgements {
This work was supported in part by Conselho Nacional de Desenvolvimento
Cient\'{\i}fico e Tecnol\'ogico - CNPq - Brazil and
by the Russian Foundation of Fundamental Investigation ( Grant 99-02-17646). }

\newpage 
%


\newpage
\Large
\begin{center}
Figure Captions
\end{center}
\normalsize
\vspace{1cm}

\noindent Figure 1 - Example of configurations of molecules with distinct 
sizes $s$ in a lattice of size $N=6$. The coordinates of the molecules  
are denoted by the black squares.

\noindent Figure 2 - Graphical representations of: a)  the $S$ matrix \rf{28}
 and b) the Yang-Baxter equations \rf{36}.

\noindent Figure 3 - Graphical representation of the relation \rf{33}. 
The circles represents the particle's indices $\{Q\}$ and $\{Q'\}$, 
and the lines the ``wave numbers" $\{k_{P_i}\}$. The variables represented in 
the full circles are summed.

\noindent Figure 4 - a) Graphical 
representation of the result of $n$ iterations 
of relation \rf{33} on the right of equation \rf{35}. The diagramatic 
representation of  Fig. 3 was used. b) A graphical representation of 
equation \rf{38}.

\noindent Figure 5 - Graphical representation of the transfer matrix $T(k)$, 
of the inhomogeneous 6-vertex model, in a periodic lattice of size $n$.

\noindent Figure 6 - Graphical representation of the relation \rf{42} 
between the $S$ matrix and the monodromy matrix ${\cal M}$.

\noindent Figure 7 - Graphical representation of  the relations \rf{46}.

\end{document}